\documentclass{sf2a-conf}
\usepackage{graphicx}
\newcommand{\zphot}{$z_{phot}$~}
\newcommand{\zphots}{$z_{phot}s$~}

\begin{document}

\TitreGlobal{SF2A 2006}

\title{Color distribution of galaxies in the CFHTLS-Deep fields}
\author{Florence Ienna, and Roser Pell\'{o}}\address{Laboratoire
d'astrophysique de Toulouse et Tarbes,UMR 5572, Universit\'{e} Paul Sabatier
Toulouse 3, CNRS, 14 avenue Edouard Belin, 31400 Toulouse, FRANCE.} 
\runningtitle{Colors of galaxies in the CFHTLS-Deep Survey}
\setcounter{page}{237}
\index{Florence Ienna and Roser Pello}

\maketitle
\begin{abstract}
We present the results obtained on the color distribution of galaxies 
in the CFHTLS-Deep Field Survey Data Release 03
\footnote[2]{Based on observations obtained with MegaPrime/MegaCam, a joint
  project of CFHT and CEA/DAPNIA, at the Canada-France-Hawaii Telescope
  (CFHT), which is operated by the National Research Council (NRC) of Canada,
  the Institut National des Sciences de l'Univers of the Centre National de la
  Recherche Scientifique (CNRS) of France, and the University of Hawaii. This
  work is based in part on data products produced at TERAPIX and the Canadian
  Astronomy Data Center as part of the Canada-France-Hawaii Telescope Legacy
  Survey, a collaborative project of NRC and CNRS.}. 
Photometric redshifts have been computed using a standard SED fitting
approach, with a new version of the public code HyperZ (New-HyperZ). 
Large samples of galaxies with well determined photometric redshifts 
in the $0<z<1.3$ interval have been selected in the four CFHTLS Deep fields,
within the completeness limit in absolute luminosity in $u$ and $r$
bands. We study the restframe color distribution of galaxies as a function of
redshift, luminosity and local density. 
Our results are consistent with a bimodal color distribution, where 
red galaxies dominate the highest luminosities out to $z \sim 0.6$. An 
important population of blue and bright galaxies appears beyond this redshift,
increasing with redshift. Out to $z \sim 1.3$, a
strong evolution is observed, at a given redshift, in the color distribution 
of galaxies as a function of luminosity, together with a mild evolution with 
the local density at fixed luminosity.  
\end{abstract}

\section{Introduction}
  Considerable progress has been made during the last ten years on the study
of galaxy properties and their evolution out to redshifts $z \sim 1$, thanks
to large surveys (e.g.:Canada-France Redshift Survey; Lilly et
al.\ 1995; 2 Degree Field Survey; Colless et al.\ 2001). 
 More recently, the SDSS (Sloan Digital Sky Survey; York et al.\ 2000) has allowed
the extragalactic comunity to study the local universe using an extremely large
sample of galaxies. 
Different attempts have been made to constrain the
global photometric properties of galaxies up to $z\sim 6$ based on deep
high-resolution imaging surveys (Hubble Deep Fields; Williams 
et al.\ 1996; Hubble Ultra Deep Field; Beckwith et al.\ 2006).
 However, these deep pencil-beam surveys based 
on photometric redshifts include a relatively small sample of galaxies
 compared to local samples. 
All this studies have demonstrated the existence of strong correlations between 
galaxy rest-frame properties, morphology and environment.
Subtantial progress will be provided by Large spectroscopic datasets, 
combined with photometric multi-band data 
(e.g.:Franzetti et al.\ (2006) using the VIMOS-VVDS data). 

  The combination of homogeneous wavelength coverage and photometric depth on 
the large effective area achieved by the CFHTLS (Canada-France-Hawaii
Telescope Legacy Survey) allows the study of galaxy evolution with 
unprecedented accuracy and is particularly well suited for a detailed
 study of galaxy populations at $z\sim0.2-1.3$.  
A recent paper by Nuijten et al. (2005) 
addresses the relationships between galaxy morphology, luminosity, color and
environment as a funtion of redshifts using a sample of $\sim$65000 
galaxies selected in one of the CFHTLS Deep fields (hereafter CFHTLSD). These authors find a bimodal
color distribution for galaxies out to $z\sim1$, with a prominent red sequence 
at $0.2<z <0.4$, and a large blue population at $0.8<z<1$.

     We analyse in this paper the color distribution of galaxies out to
$z\sim1.3$, using a template fitting method to derive photometric
redshifts and restframe properties for galaxies in the T0003 release
of the CFHTLSD. In Sect.\ \ref{data} we describe the CFHTLSD data used 
in this study.  Photometric redshifts are presented in Sect.\ 
\ref{photoz}, toghether with the sample selection used in this study.
Results are summarized in Sect.\ \ref{results}.
More details will be given in a forthcoming paper (Ienna \& Pell\'o, in
preparation). We adopt the cosmological parameters
$\Omega_{\Lambda}=0.7$, $\Omega_{m}=0.3$, and $H_{0}=70\ km\ s^{-1}\ 
Mpc^{-1}$. Magnitudes are given in the AB system (Oke 1974).

\section{\label{data} The Data}

   The CFHTLSD covers 4 square degrees in four independent fields (D1, D2, D3,
D4), observed through five bands in the visible domain: $u^{\ast}, g', r', i',
z'$. The data were processed at Terapix (astrometry, photometric calibration,
image stacking, detection and photometric catalogs). 
More details can be found on the
Terapix\footnote{http://terapix.iap.fr} webside.  
In this study we used the CFHTLSD T0003 release wich includes observations
from June 2003 to July 2005. The image quality in the final stack ranges
between 0.8$''$ in $z'$ and 1.0$''$ in $u^{\ast}$. 
The four catalogs released by Terapix included
more than 1.6 million objects in total, up to $AB\sim$ 27.3 in $u^{\ast} g'$,
$AB\sim$ 27 in $r' i'$, and $AB\sim$ 26 in $z'$
(SExtractor MAG\_AUTO, 1$\sigma$ detection limit). 

\section{\label{photoz} Photometric redshift and sample selection}

\begin{figure}[h]
\includegraphics[width=.7\textwidth,angle=270]{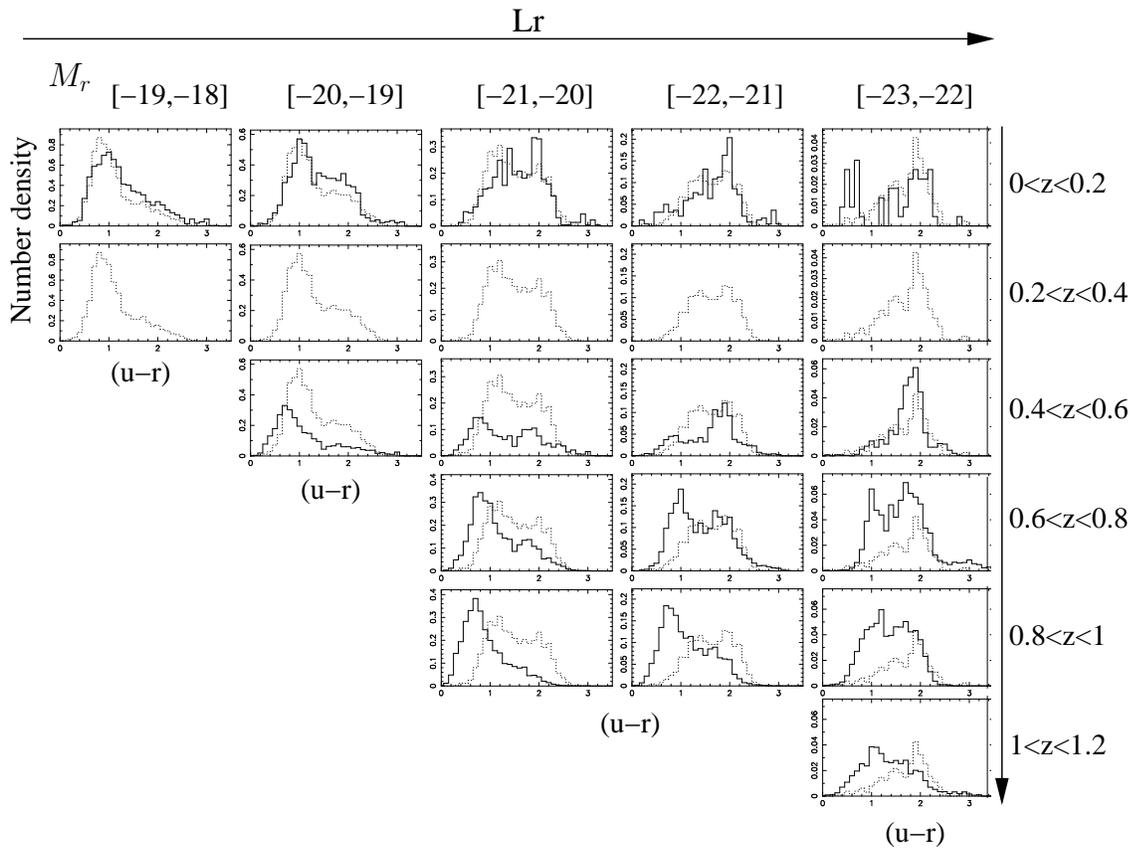}
\caption{Number density of galaxies (in units of 10$^{-3}$ galaxies Mpc$^{-3}$)
as a function color. The evolution of this relationship is presented as a 
function of redshift (increasing from top to bottom) and luminosity
(increasing from left to right). The distribution of galaxies in the z=0.2-0.4
bin is superimposed to all diagrams for comparison (dashed line). 
The histogram at the bottom left presents the 
redshift distribution for different i-band selected samples. This
diagram is a combination of the four CFHTLS Deep Fields.
}
\label{fig1}
\end{figure}
 
  Photometric redshifts (hereafter \zphot ) have been
computed with a new version of the public code {\it Hyperz\/} ({\it
new\_hyperz\/}
\footnote{http://www.ast.obs-mip.fr/users\/roser/hyperz/}
), originally developped by Bolzonella et al. (2000).
This method is based on the fitting of the photometric Spectral Energy
Distributions (SED) of sources using a large set of templates. 
 In this case, the template library includes 14 templates: 8 evolutionary 
synthetic SEDs computed with the last version of the Bruzual \& Charlot 
code (Bruzual \& Charlot 1993), with Chabrier (2003) IMF;
4 empirical SEDs from Coleman, Wu and Weedman (1980), and 2 starburst galaxies
 from Kinney et al.\ (1996).Internal extinction is considered as a free parameter following the
Calzetti's (2000) law, with E(B-V)$\sim$0-0.45 mags.
Galactic extinction is also corrected. 
Photometric redshifts were computed in the range z$=$0-6.
No luminosity prior was used, but a simple cut in the permitted range of
luminosities: M$_B$=$[$-14,-23$]$.
SExtractor MAG\_AUTO magnitudes and errors were used to compute \zphots. 
When an object is non-detected in a given filter, the flux in this filter is
set to 0, with an error bar corresponding to a S/N$\sim$1 in this filter.
We have also corrected for seeing differences between the different images, 
taking the i-band image as a reference.
 The photometric redshift accuracy was estimated by a direct
comparison with secure spectroscopic samples available in the Groth/Deep
Survey (D3) (see more details in our web page\footnote{http://www.ast.obs-mip.fr/users/roser/CFHTLS\_T0003/}
). Similar results are obtained using the VVDS spectroscopic sample (D1, Ilbert
et al.\ 2006). Due to the lack of near-IR filters, we expect to obtain
accurate \zphots for our sample up to $z\sim1.3$. 
A blind comparison with 
spectroscopic redshifts yields a mean dispersion $\sigma(\Delta
z/(1+z))$=0.056 for the whole sample within $z=0-1.3$.
All objects in masks, saturated sources and bright stars up to $i<22$ have
been removed from the sample. 
Only sources detected in at least two different
filters with S/N$\ge$3 have been considered. 
An additional selection has been introduced based on the quality of the fit,
affecting only $\sim$10\% of the remaining objects.
The final sample used in this study contains 
1.13 million sources over 4 square degrees.

\section{\label{results} Results: Color distribution out to z$=$1.3}

\begin{figure}[h]
\includegraphics[width=.6\textwidth,angle=0]{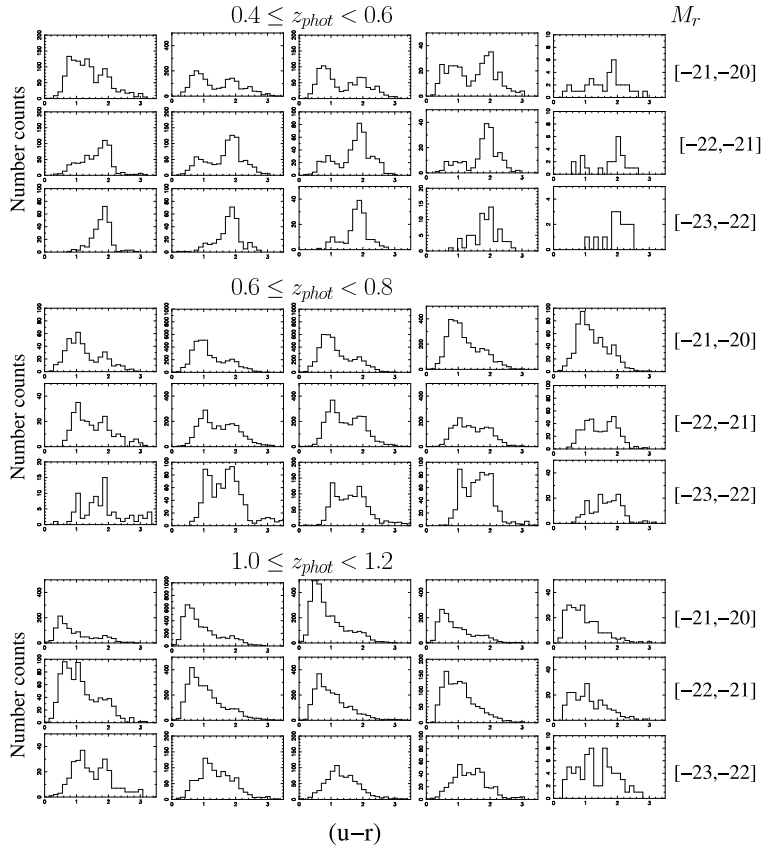}
\caption{Number density of galaxies (in units of $10^3$ galaxies Gpc$^{-3}$)
as a function color. The evolution of this relationship is presented as a function
of the local density (increasing from left to right) and luminosity, for
three representative redshifts.
}
\label{fig2}
\end{figure}

 In order to address the evolution of the color distribution of galaxies as a
function of redshift and local density, we have defined samples of sources
which are complete in {\it all} the filters considered simultaneously 
(here the restframe $u^{\ast}$ and $r'$ bands). 
Different samples are selected here in absolute magnitude M$_r$ as a function of
redshift, in such a way that for galaxies brighter than M$_r$, the limiting
M$_u$ allows to reach at least a restframe $u-r \le 3.5$.
The limiting magnitudes in this complete sample of $\sim$0.9 million galaxies
are M$_r \le$ -18 for $z\le$0.4, 
M$_r \le$ -19 for $z\le$0.6, M$_r \le$ -20 for $z\le$1.0, and M$_r \le$ -22 for
$z\le$1.2. 
The evolution of the restframe color $u-r$ as a function of redshift and
luminosity is presented in Fig.~\ref{fig1}. 
A bimodal color distribution is observed at
all redshifts and luminosities. From recent studies of the local universe
(e.g. Balogh et al. 2004, Hogg et al. 2003, Cooper et al. 2006), it is
admitted that this bimodal distribution corresponds to a separation between
red early type galaxies (ellipticals with old stellar passively evolving
population) and blue late type galaxies (spirals with a young stellar
population). 
The color distribution shown in Fig.~\ref{fig1} is bimodal out to $z\sim1.2$.
The red population is found to dominate the brightest luminosity bins out to
$z\sim0.6$. Its mean color becomes bluer with increasing redshift. Although a
red and bright population still exists at z$\sim1.2$, the blue population 
dominates at z$\sim$0.8-1.2 in all the luminosity bins.
The blue population also dominates in the lowest luminosity bins at low
redshift, and becomes more important with increasing redshift. Also the mean
color of this population becomes bluer with increasing redshift.  
Our results at z$\le0.4$ are in good agreement with those found in the local
universe (e.g. Baldry et al.\ 2004), although 
our sample is much less significant in the brightest bin at z$\le0.2$. 

We have studied the relationship between color distribution and environment
taking advantage of the large complete sample of galaxies in the CFHTLSD.  
A local density estimator has been derived for each object based  
on the distance to the 10th closest neighbour D: $Sigma_{10}= 10/(\pi*D^2)$,
where D is the projected linear distance at \zphot. Close neighbours are
selected within a photometric redshift slice \zphot$\pm0.1$, with M$_r$ in the
[-24,-20] interval. This is the \zphot equivalent of the estimator used by
Balogh et al.\ (2004) with the SDDS spectroscopic sample. 
Fig.~\ref{fig2} displays the
number density of galaxies as a function color, for three representative
redshift bins, and for five density regimes (roughly the same used by Balogh
et al.\ (2004) in the SDSS). At redshift between $0<z<0.6$ we observe the same
trends as in the SDSS. The blue population dominates the lowest density and
faintest luminosity bins, whereas the red population is found to dominate the
highest density and brightest luminosity bins. 
These general trends seem more sensitive to the luminosity than to the local
density. At $z>0.8$, a bright blue population appears even in the densest
regions, increasing with redshift. 

  In conclusion, the optimal combination of homogeneous wavelength coverage
and photometric depth on the large effective area achieved by the CFHTLSD T0003
allows the study of photometrical properties of galaxies with unprecedented
accuracy. 



\begin{thebibliography}{}
\bibitem{}Balogh, M. L., Baldry, I.K., Nichol, R.,Miller,C. ,Bower, R. and Glazebrook, K.\. 2004, ApJ,615, L101 
\bibitem{}Baldry, I.K., Balogh, M. L., Bower, R., Glazebrook, K. and Nichol, R.\. 2004, AIP Conf. Proc., 743(2004) 100-119
\bibitem{}Beckwith, S.~V.~W., et al.\ 2006, AJ, 132, 1729 
\bibitem{}Bolzonella, M., Miralles, J. M., \& Pello, R.,\ 2000, A\&A, 363, 476-492
\bibitem{}Chabrier, G.\ 2003, PASP, 115, 763 
\bibitem{}Coe, D., Ben{\'{\i}}tez, N., S{\'a}nchez, S.~F., Jee, M., Bouwens,  R., \& Ford, H.\ 2006, AJ, 132, 926  
\bibitem{} Coleman, G.~D., Wu, C.-C., \& Weedman, D.~W.\ 1980, ApJ, 43, 393 
\bibitem{}Colless, M.M.,et al.\ 2001, MNRAS, 328, 1039
\bibitem{}Cooper, M. C., et al.\ 2006, MNRAS, accepted [astro-ph/0603177]
\bibitem{}Croton, D.~J., et al.\ 2005, MNRAS, 356, 1155 
\bibitem{}Dressler, A. \ 1980, ApJ,236,351
\bibitem{}Franzetti, P, et al.\ 2006, submitted to A\&A, astro-ph/0607075.
\bibitem{}Hogg, D. W., et al.\ 2003, ApJ, 585, L5
\bibitem{}Ilbert, O., et al.\ 2006, A\&A, 457, 841 
\bibitem{}Kinney et al. \ 1996
\bibitem{}Kennicut, R. C., Jr.\ 1983, AJ, 88, 483
\bibitem{}Lilly, S. J., Le f\`{e}vre, O., Crampton, D., Hammer, F,\& Tresse, L.\ 1995, ApJ, 455, 50
\bibitem{}Nuijten, M.~J.~H.~M., Simard, L., Gwyn, S., Rottgering,  H.~J.~A.\ 2005, ApJ, 626, L77  
\bibitem{}Wiegert, T., de Mello, D.~F., \& Horellou, C.\ 2004,  A\&A, 426, 455 
\bibitem{}Williams, R. E., et al.\ 1996, AJ, 112, 1335
\bibitem{}Withmore, B. C., Gilmore,D. M.,\& Jones, C.\ 1993, ApJ, 407, 489
\bibitem{}York, D. G.,Adelman, J., Anderson, J. E. et al.\ 2000, AJ, 120, 1579
\end{thebibliography}
\end{document}